\def\titre#1{\bigbreak{\bf#1}\par\penalty100\medskip}
\def\titr#1{\medbreak{\bf#1}\par\penalty100\medskip}

\def\ent#1{{\left\lfloor#1\right\rfloor}}

\def\abs#1{{\left\vert#1\right\vert}}
\let\ds=\displaystyle
\begingroup
\catcode`\{=11 \global\let\lacco={
\catcode`\}=11 \global\let\racco=}
\endgroup
{\def\aaaa{\catcode`\\=11 \global\let\bksl} \aaaa=\ }
{\obeyspaces\gdef\blanc{\obeyspaces\let =\ }}
\let\qar=\par
\def\ligne{\qar\hskip0pt}
\let\dblq="
{\catcode`\"=\active
\gdef\debchaine{\dblq\catcode`\\=11 \let"=\finchaine}
\gdef\finchaine{\dblq\catcode`\\=0  \let"=\debchaine}
\global\let"=\debchaine
}
\def\tty#1{\let\suite=\tty
\ifx#1.\let\suite=\tt \parindent=0pt \obeylines \blanc \else
\ifx#1;\let\suite=\relax\else
\ifx#1c\def\suite{\begingroup\def\n{\bksl n}\tty p"~\%&\{\}}\else
\ifx#1n\def\suite{\tty \\$}\let\{=\lacco \let\}=\racco \else
\ifx#1p\def\suite{\tty ^_\#}\let\\=\bksl\else
\ifx#1"\catcode`"=\active\else\catcode`#1=11
\fi\fi\fi\fi\fi\fi\suite}
\def\verbatim{\tty p&.}
\def\verbatimn{\tty pn.}
\def\verbatimc{\tty c.}
\def\verbatimcinput#1{\verbatimc\let\par\ligne\input#1\endgroup}
\def\verbatimninput#1{{\verbatimn\let\par\ligne\input#1}}
\def\br{\hfill\break}
\def\pmod#1{\;({\rm mod}\,#1)}
\def\n{{\it n}}
\titre
{Fortuitous sequences of flips of the top of a stack of $n$ burnt pancakes for all $n>24$.}
\medskip
Laurent PIERRE, universit\'e de Nanterre, France

{\tt Laurent.PIERRE@parisnanterre.fr}

\titre{I Introduction}
In 1979, William H. Gates (aka Bill Gates) and Christos H. Papadimitriou
wondered how many spatula flips are needed to reverse a stack of burnt pancakes.
They proved that $g(-f_n)\ge 3n/2-1$.

In 1995, Cohen and Blum conjectured that $g(n)=g(-I_n)$, i.e.
the hardest case is when pancakes are upside down in right order.
They proved that $g(-I_n)=g(-f_n)+1\ge 3n/2$.

In 1997, Heydari and Sudborough proved that $g(-I_n)\le 3{n+1\over 2}$ whenever some ``fortuitous sequence'' of flips exists.
They gave fortuitous sequences for $n=3$, 15, 23, 27 and 31.
They showed that two fortuitous sequences $S_n$ and $S_{n'}$ may combine into another fortuitous sequence $S_{n''}$
with $n''=n+n'-3$.  So a fortuitous sequence $S_n$ gives a fortuitous sequence $S_{n+12}$.
This proves that $g(-I_n)\le 3{n+1\over 2}$ if $n\equiv 3 \pmod 4$ and $n\ge 23$.

In 2011, Josef Cibulka used a potential function and obtained a better lower bound : $g(-I_n)\ge 3n/2+1$ if $n>1$,
proving thereby that $g(-I_n)=3{n+1\over 2}$ if $n\equiv 3 \pmod 4$ and $n\ge 23$.
Cibulka gave also a sequence of 30 flips
(19, 14, 7, 4, 10, 18, 6, 4, 10, 19, 14, 4, 9, 11, 8, 18, 8, 11, 9, 4, 14, 19, 10, 4, 6, 18, 10, 4, 7, 14)
sorting $-I_{19}$, without being fortuitous.
With the fortuitous sequence sorting $-I_{15}$, this proved that
$g(-I_n)=3{n+1\over 2}$ if $n\equiv 3 \pmod 4$ and $n\ge 15$.
Cibulka also invalidated Cohen and Blum's conjecture by showing that $g(-I_{15})=24$ while $g(15)=25$.

This paper gives $g(-I_n)$ for every $n$
\halign{&~\hss$#$\hss~\cr
n       &1&2\ldots7&  8&  9& 10& 11& 12& 13& 14& 15& 16& 17& 18& 19& 20& 21& 22& 23\ldots\cr
g(-I_n) &1&2n      & 15& 17& 18& 19& 21& 22& 23& 24& 26& 28& 29& 30& 32& 33& 35& \left\lceil{3\over 2}n\right\rceil+1\cr}
and explains how to build for every $n\ge25$ a generalized fortuitous sequence
proving $g(-I_n)=\left\lceil{3\over 2}n\right\rceil+1$.

After this introduction, section II defines the burnt pancakes problem and $g(-I_n)$.
For $n>1$ it gives some simple bounds such as $3n/2\le g(-I_n)\le 2n$ and $g(-I_{n+1})\le 2+g(-I_n)$.
Then it proves Cibulka's bound : $g(-I_n)\ge 3n/2+1$ and characterizes the flips sequences for which this bound is tight.

Section III defines a (regular) odd fortuitous sequence almost like Heydari and Sudborough
and the combination of two such fortuitous sequences with odd $n$ and $n'$ into a fortuitous sequence for $n+n'-3$ pancakes.
It provides odd fortuitous sequences for $n=3$, 15, 23, 29, 31, 33 $\ldots$ , some of them being triple or palindromic.

Section IV defines an even fortuitous sequence which contains only 2 flips $f_n$ of the whole stack
and in which the stack of 2-clans and the stack of 2-blocks are replaced by a patchwork of a mixture of 2-blocks and 2-clans.
The combination of two fortuitous sequences with even $n$ and $n'$ is a fortuitous sequence for $n+n'-2$ pancakes.
Even fortuitous sequences are provided for $n=26$, 28, 30, 32 $\ldots$ , some of them being double or palindromic.

Section V defines generalized odd fortuitous sequences with patchworks, which combine after a regular fortuituous sequence.
It provides such sequences for $n=23$, 25, 27, 29 $\ldots$ .

Section VI gives $g(-I_n)$ for $n=24$, 21, 22 and 20, when no (generalized) fortuitous sequence exists.

Section VII explains how to build most of the (fortuitous) sequences given in this article,
with the program in C of section IX.  

Section VIII is a bibliography.

\titre{II The burnt pancakes problem : bounds on $g(-I_n)$}
The stack of burnt pancakes $\left[\vcenter{\halign{\hss#\cr2\cr-4\cr5\cr1\cr-3\cr}}\right]$
holds 5 pancakes numbered from 1 to 5 from the smallest to the largest.
In this stack pancakes 3 and 4 are upside down.
In order to save space, it will be rather written $S$=[2 -4 5 1 -3], top of stack on the left.

This stack of pancakes may be seen as a signed permutation of integers 1 to 5,
i.e. mapping $1\mapsto2$, $2\mapsto-4$,  $3\mapsto5$,  $4\mapsto1$,  $5\mapsto-3$.
This mapping may be extended to negative integers -5 to -1 by $S(-i)=-S(i)$.
Such mappings may be composed. E.g. [2 -4 5 1 -3]$\circ$[1 2 -3 5 4]=[2 -4 -5 -3 1].

Stack $I_n$=[1 2 3 $\ldots$ \n] denotes identity mapping. It is also the target of the pancakes problem.
Pancakes are sorted by size, the smallest one on top. They are all well oriented i.e. burnt side down.

Stack $-I_n$=[-1 -2 -3 $\ldots$  -\n] is one of the hardest stacks to sort for pancakes problem.
Pancakes are all in right location but they are upside down, i.e. burnt side up.

Spatula flip $f_i$ splits a stack of pancakes in two parts by inserting a spatula below the $i$ topmost pancakes,
then flips the upper part, and put it back on top of the lower part.
E.g. $f_3$=[-3 -2 -1 4 5] when the stack holds 5 pancakes.
Flips are the only operations allowed to sort the stacks of pancakes.

Any signed permutation $S$ of a stack of $n$ burnt pancakes, may be achieved by a flip sequence.
Let $g(S)$ be the minimal number of flips needed. 
Cibulka proved in 2011 that $g(-I_n)\ge 3n/2+1$ if $n>1$.

Flip sequence $w=(f_4\;f_8\;f_6\;f_{10})$ will be shortened as $w=($4 8 6 10$)$.
Let $\tilde w=($10 6 8 4$)$ and $f_w=f_{10}\circ f_6\circ f_8\circ f_4$.
Permutation $f_w$ is the effect of sequence $w$.
Applying flip sequence $S$ to stack $-I_n$ gives stack $f_S\circ{-}I_n=-f_S$.
So ``sequence $S$ sorts stack $-I_n$'' translates into $f_S\circ{-}I_n=I_n$ i.e. $f_S=-I_n$.

\titr{IIa Lower bound on $g(-I_n)$ by Cohen and Blum : $3n/2$}
When Gates and Papadimitriou introduced the burnt pancake problem in 1979,
pancakes in initial setting had right orientation,
so every single pancake had to be flipped an even number of times.
That is why, they looked at $g(-f_n)$ instead of $g(-I_n)$ : at the start all the pancakes have right orientation,
but they are sorted in reversed order $-f_n$=[\n~\n-1 $\ldots$ 3 2 1].
They proved that $g(-f_n)\ge 3n/2-1$.

In 1995, Cohen and Blum noticed that every signed permutation $p$ of a stack of $n$ pancakes verifies
$p\circ -I_n=-I_n\circ p=-p$. Hence $-I_n$ commutes to every signed permutation. (In other words, $-I_n$ is in the center
of the group of signed permutations. Indeed this center is $\{I_n,-I_n\}$).

If $-I_n=p\circ q$ then $-I_n$ commutes to $p$ and $p^{-1}$ and so $-I_n=q\circ p$.
This means that a flip sequence which sorts $-I_n$, may be permuted circularly. It will still sort $-I_n$.

Similarly, since $-I_n$ and flips are their own inverses, when a flip sequence sorting $-I_n$ is read backwards,
it still sorts $-I_n$. In other words, $f_{\tilde S}=f_S^{-1}$ so if $f_S=-I_n$ then $f_{\tilde S}=-I_n$.

For $n>1$, a flip sequence sorting $-I_n$ holds at least two $f_n$, the first one removes -\n~from the bottom of the stack,
while the last one brings $n$ back at the bottom.
Hence any minimal sequence sorting $-I_n$ may be permuted circularly to produce another minimal sequence starting by $f_n$.
Removing this initial $f_n$ yields a minimal sequence sorting $-f_n$.
Hence $g(-I_n)=1+g(-f_n)$.
This is how Cohen et al. proved that $g(-I_n)\ge 3n/2$.

\titr{IIb Simple upper bounds for $g(-I_n)$ : $g(-I_n)\le2n$ and $g(-I_{n+1})\le 2+g(-I_n)$.}
$f_{n-1}\circ f_n=[$2 3 $\ldots n$-1 ~$n$~ -1] permutes circularly the pancakes and flips one of them.
So $(f_{n-1}\circ f_n)^n=-I_n$.
Sequence $(n~n{-}1)^n$ sorts stack $-I_n$. Hence $g(-I_n)\le2n$.
Indeed $g(-I_n)=2n$ ~ if ~ $1<n\le7$.

Actions of (2 1$)^2$ on -$I_2$ and (3 2$)^3$ on -$I_3$ can be compared :
\halign{&\hss#\hss&~$\to$~#\cr
[-1 ~-2~~]&&[~2~ 1]&&[~-2~~ 1]&&[-1 ~2~]&\omit&   &&    &\omit&[1 ~2~]\cr
[-1 -2 -3]&&[3 2 1]&&[-2 -3 1]&&[-1 3 2]&&[-3 1 2]&&[-2 -1 3]&&[1 2 3]\cr}
In the first part we replace 2 by 3~2 (and -2 by -2~-3).
In the  last part we replace 2 by 2~3 (and -2 by -3~-2).
In between, we replace $f_1$ by $f_2$ $f_3$ $f_2$.
This generalizes as follows :
Let $S$ be a sequence of minimal length sorting stack $-I_n$.
Let $f_i$ be the first flip of this sequence. If $i=n>1$ we can remove this first flip of $S$ and put it at the end.
So we make sure that $i\ne n$. In the stack we replace pancake $i+1$ by two pancakes. In sequence $S$ 
a flip $f_j$ becomes $f_{j+1}$ if the doubled pancake belongs to the flipped upper part of the stack.
So sequence $S$ becomes sequence $S'$,
which sorts stack [-1 $\ldots$ -$i$ $|$ -$i$-2 -$i$-1 $|$ -$i$-3 $\ldots$ -$n$-1$]=-f_{i+1}f_{i+2}f_{i+1}f_i$.
The first flip of $S'$ is $f_i$. Replacing it by sequence $f_{i+1}$ $f_{i+2}$ $f_{i+1}$ yields
sequence $S''$ of length $g(-I_n)+2$ sorting $-I_{n+1}$.
This proves that $g(-I_{n+1})\le 2+g(-I_n)$ ~if~ $n>1$.

\titr{IIc Lower bound for $g(-I_n)$ with a potential function}
In a stack of pancakes a ``block'' is a maximal sequence of several contiguous pancakes which appears also in $I_n$ or $f_n$.
A ``clan'' is a maximal sequence of several contiguous pancakes which appears also in $-I_n$ or $-f_n$.
Two contiguous pancakes in a same block are ``adjacent''.
Two contiguous pancakes in a same clan are ``anti-adjacent''.
A pancake which is neither in a block nor in a clan is a ``singleton''.
E.g. stack [4 5 6 -1 -2 3] is made of block 4 5 6, clan -1 -2 and singleton 3.
The potential of a pancake in a stack will be defined as 1 when at an end of a block, 3 when inside a block,
-1 when at an end of a clan, -3 when inside a clan and 0 for a singleton.
The potential of the top pancake is doubled.
The potential of a stack will be defined as the sum of the potentials of its pancakes. E.g. the potential of stack
$\vcenter{\halign{&$\;$#\cr
[4&5&6&-1&-2&3]\cr\hss
 2&3&1&-1&-1&0\hss\cr}}$
is 4.

\def\tto{{\rightarrow}}
\def\ttot{{\leftrightarrow}}
During the flip $[-x\ldots y|z\ldots]\to[-y\ldots x|z\ldots]$, the potential function varies only for pancakes $x$, $y$ and $z$.
The possible variations of $p(y)$ are $3\tto2$, $1\tto2$, $1\tto0$, $0\tto0$, ${-}1\tto0$, ${-}1\tto{-}2$ and ${-}3\tto{-}2$.
Hence $|\Delta p(y)|\le 1$.  As well $|\Delta p(x)|\le 1$.
The possible changes of $p(z)$ are $3\ttot1$, $1\ttot0$, $0\ttot{-}1$, $1\ttot{-}1$ and ${-}1\ttot{-}3$.
Hence $|\Delta p(z)|\le 2$.
Hence $|\Delta p(x)+\Delta p(y)+\Delta p(z)|\le 1+1+2=4$.

During the flip $[-x|z\ldots]\to [x|z\ldots]$ the potential function varies only for pancakes $x$ and $z$.
The possible variations of $p(x)$ are 2$\ttot0$, $0\tto0$ and $-2\ttot0$. Hence $|\Delta p(x)|\le 2$.
We still have $|\Delta p(z)|\le 2$.
Hence $|\Delta p(x)+\Delta p(z)|\le 4$.
This proves that during a flip, the potential function increases by at most 4.

$p(I_n)=2+3+3+\cdots+3+1=3n-3$.\qquad
$p(-I_n)=p(-f_n)=-p(I_n)=3-3n$.

Hence\qquad $\ds g(-f_n)\ge{p(I_n)-p(-f_n)\over4}={3n-3\over2}$\qquad and\qquad $\ds g(-I_n)\ge{3n-1\over2}$.

\titr{IId Enhancement of potential function to reach Cibulka's bound : $g(-I_n)\ge 3n/2+1$}
The potential function may be enhanced if the plate supporting the stack behaves like a pancake,
which may form a block with the bottom pancake when it is $n$ or a clan if it is 1.
The plate potential is -1, 0 or 1 and adds to the potential function of the stack.
The final potential is now $p(I_n)=3n$ and the initial potential is $p(-f_n)=-3n$.
Then $g(-f_n)\ge 6n/4=3n/2$.
Hence $g(-I_n)\ge 3n/2+1$.
{\verbatim
   [3 2 1]     [-2-3 1]     [-1 3 2]     [-3 1 2]     [-2-1 3]      [1 2 3]
-9=-2-3-3-1  -5=-2-1-1-1   -2=0-1-1      2=0+1+1      5=2+1+1+1    9=2+3+3+1
           +4           +3           +4           +3            +4         
}
Since $g(-I_n)$ is an integer, the lower bound is actually : $g(-I_n)\ge \left\lceil{3n\over 2}+1\right\rceil=\left\lfloor{3n+3\over 2}\right\rfloor$.

This was proved in 2011 by Cibulka who used about the same potential function with the following differences :

* His potential function is the third of ours. E.g. potential function variation during a flip is at most 4/3.

* The potential of a block with $n_a$ adjacencies is $3n_a$ if it is on top of the stack
and $3n_a-1$ if it is inside the stack.
As well the potential of a clan with $n'_a$ anti-adjacencies is $-3n'_a$ when on top of the stack and $-3n'_a+1$ otherwise.
Hence the potential of a stack $S$ is $p(S)=p'(S)-p'(-S)$ with $p'(S)=3n_a(S)-n_b(S)$,
where $n_a$ and $n_b$ count adjacencies and inner blocks.
Cibulka takes $p(S)=n_a(S)-n_a(-S)-(n_b(S)-n_b(-S))/3$.
 
* Cibulka keeps a neutral plate under the stack of pancakes, but he adds to the potential 4/3, 1, -1 or -4/3
if the bottom of stack is ~\n-1,\n~ or ~\n~ or ~-\n~ or ~1-\n,-\n~ respectively.

* Cibulka starts from $-I_n$ instead of $-f_n$, but he adds 1/3 to the potential when a block holds 1,2 (or -2,-1) or
when -1 is on top of the stack.
As well he adds -1/3 to the potential when a clan holds 2,1 (or -1,-2) or when 1 is on top of the stack.
 
\titr{IIe Flip sequence reaching the lower bound}
Let us assume $g(-I_n)=\left\lceil 3n/2+1\right\rceil$. If $n$ is even, then $g(-f_n)=3n/2$.
There exists a flip sequence of length $3n/2$ transforming $-f_n$ into $I_n$.
The potential function goes from $-3n$ to $3n$ in $3n/2$ flips. Hence every flip increases potential by 4.
For every flip $f_1$ we have $\Delta p(x)=\Delta p(z)=2$.
For every larger flip we have $\Delta p(x)=\Delta p(y)=1$ and $\Delta p(z)=2$.
The potential function of a pancake increases by 1 when it leaves or reaches the top of the stack,
and increases by 2 when not moved.
This proves that the parity of the potential of a pancake depends only on its location in the stack :
Every pancake but top one has an odd potential.
Only top pancake may be a singleton.
The plate potential starts from -1 in $-f_n$ and ends at 1 in $I_n$. It increases by 2 during every $f_n$.
So only one $f_n$ is used to transform $-f_n$ into $I_n$, and two of them are used to transform $-I_n$ into $I_n$.

If $n$ is odd, then the lower bound is reached when $g(-I_n)=(3n+3)/2$.
There exists a flip sequence of length $(3n+1)/2$ transforming $-f_n$ into $I_n$.
The potential function goes from $-3n$ to $3n$ in $(3n+1)/2$ flips.
Its variation is loose by 2.
There may be two $f_n$. Then the first $f_n$ changes the plate potential from -1 to 0 and the second $f_n$ from 0 to 1.
The variation of the potential function during both $f_n$ will be 3. Hence its variation during every smaller flip will be 4.
The transformation from $-I_n$ to $I_n$ uses three $f_n$.

We shall build flip sequences reaching the lower bound for every $n$ large enough.
\vfill\eject
\titre{III Regular (odd) fortuitous sequences and stacks of 2-clans}
A ``fortuitous sequence'' is a flip sequence of the form $S=ns_1ns_2ns_3$ where $s_1$, $s_2$ and $s_3$ are
sequences of length ${n-1\over2}$ of even positive integers lower than $n$, such that $f_S=-I_n$.

So $n$ is odd and $\abs S={3n+3\over 2}$. The lower bound is reached.
Hence when applying $S$ to $-I_n$, sequence $s_1ns_2ns_3$ transforms $-f_n$ into $I_n$.
Both $f_n$ increase the potential function by 3
and all the other flips increase it by 4.
Since $s_1$ is made of even numbers, when applying $s_1$ to $-f_n$, pancake 1 stays at the bottom of the stack,
and above it, is a stack of clans of even lengths.
Each flip, to increase the potential function by 4, must break a clan into two smaller clans.
At the end of $s_1$ the stack $f=-f_{ns_1}$ is a stack of ${n-1\over2}$ clans of 2 pancakes placed on top of singleton 1.
Heydari et al. called it the first column of a box.
We shall call it a ``stack of 2-clans''.
Similarly every flip of $s_3$ glues two blocks to form a bigger one.
And stack $-f_{ns_1ns_2n}$ at the beginning of $s_3$ is a stack of ${n-1\over2}$ blocks of 2 pancakes,
placed on top of singleton $n$.
Hence every flip of $s_2$ replaces a 2-clan by a 2-block.
The first flip of $s_2$ is $[-1\ldots3|2\ldots]\to[-3\ldots1|2\ldots]$.
The second flip of $s_2$ is $[-3\ldots5|4\ldots]\to[-5\ldots3|4\ldots]$.
The third flip of $s_2$ is $[-5\ldots7|6\ldots]\to[-7\ldots5|6\ldots]$. Etc..
So during flip sequence $s_2$, the top pancake becomes successively -1, -3, -5, -7, $\ldots$, $-n$.
The bottom pancake does not move, but it leaves its initial clan to join its final block during the only $f_{n-1}$.
This proves that $s_2$ holds exactly one occurrence of $n{-}1$ and the bottom pancake is even and positive
i.e. right-side up.
So the top pancake of stack of 2-clans $f$ is even and negative i.e. upside down.
Of course $s_1$ holds also exactly one occurrence of $n{-}1$, and $s_3$ too.

For Heydari et al., the top pancake in the stack of 2-clans should be $-{n+1\over 2}$.
But for us, it is only even and negative.
So for them, $n\equiv3\pmod4$, but for us $n$ need only be odd.

We may notice that $h=f_n\circ f^{-1}\circ f_n$ is the stack of 2-clans of
fortuitous sequence $n\tilde s_1n\tilde s_3n\tilde s_2$ and
that $s_1$, $\tilde s_1$, $s_2$ and $\tilde s_3$ are
the only flip sequences adding an adjacency at each step when starting respectively from stacks
$-h$ (or $-f_n\circ f^{-1}$),
$-f$,
$h^{-1}$ (or $f_n\circ f$) and
$f^{-1}$ (or $f_n\circ h$).
So stack of 2-clans $f$ characterizes flip sequence $S$.

We may roughly count the fortuitous sequences.
The first flip of $s_1$ is a number taken among 2, 4, 6, $\ldots$  \n-1. There are ${n-1\over2}$ possibilities.
The second flip must break a clan. There are ${n-1\over2}-1$ possibilities.
The third flip is to be chosen among ${n-1\over2}-2$ possibilities. Etc..
So there are ${n-1\over2}!$ ways of choosing $s_1$ to get a stack of 2-clans.
After that $s_2$ and $s_3$ are imposed.
Indeed there is only one way to create an adjacency by a flip.
This flip has about one chance out of two to exist, whether the pancake matching the first one shows its right side or not.
So the expected number of fortuitous sequences may be estimated as
${n-1\over2}!/2^{n-1}\sim\sqrt{\pi n}\left({n-1\over8e}\right)^{n-1\over2}\to\infty$.
This number is large as soon as $n>8e+3$.
This suggests, that fortuitous sequences exist for every large enough odd integer $n$.
This can be proved in two different ways, that we shall both use.
Either we look for regular patterns in stacks of 2-clans of small fortuitous sequences and
design according to these patterns a stack of 2-clans yielding a fortuitous sequence for any large odd integer $n$.
Or we exhibit small fortuitous sequences with symmetries, like palindromic or triple fortuitous sequences
and use method of Heidari et al. to combine them into larger fortuitous sequences.

\titr{IIIa Combination of fortuitous sequences}
Two fortuitous sequences for $n_1$ and $n_2$ pancakes may combine into a fortuitous sequence for $n_1+n_2-3$ pancakes.
E.g. fortuitous sequences\br
(15 10  4  6 14  6  4 10 15 10  4  6 14  6  4 10 15 10  4  6 14  6  4 10)\br
(23 14  4  6 22 10  8 12 10 14 12 18 23 10 14 18  8 10 22 10  8 18 14 10 23 18 12 14 10 12  8 10 22  6  4 14)\br
yield stacks of 2-clans :\br
[-8 -9 -12 -13 -2 -3 -14 -15  5  4  7  6 -10 -11$|$1]\br
[-12 -13$|$-16 -17 -6 -7 -18 -19 -4 -5 -20 -21 -2 -3 -22 -23  9  8 11 10 -14 -15  1]\br
In the second stack, the first 2-clan -12(=-$x$) -13 is replaced by the first stack without its final 1.
Also in the first stack, pancakes 2$\ldots$ 15(=$n_1$) are renumbered 12(=$x$)$\ldots$ 25
by increasing their numbers by 10(=$x-2$).
Similarly in the second stack, pancakes 14(=$x+2$)$\ldots$ 23(=$n_2$) are renumbered 26$\ldots$ 35(=$n_1+n_2-3$)
by increasing their numbers by 12(=$n_1-3$).
This yields stack of 2-clans\br
[-18 -19 -22 -23 -12 -13 -24 -25 15 14 17 16 -20-21$|$-28-29 -6 -7-30-31 -4 -5-32-33 -2 -3-34-35  9  8 11 10-26-27  1]\br
and fortuitous sequence\br
(35 26 16 10  4  6 18 34 10  8 12 10 14 12 30  6  4 10 35 10 14 18  8 10 30  4  6 34 26 24 30 16  8 24 20 16 35 26 20 22 30  6  4 26 12 14 10 12  8 10 34 18 16 26)

Since there is a fortuitous sequence for 15 pancakes, any fortuitous sequence for $n$ pancakes yields fortuitous sequences
for $n+12$, $n+24$, $n+36$, etc..
Moreover a fortuitous sequence for $n$ pancakes with gcd($n{-}3$,12)=2 will yield fortuitous sequences for $m$ pancakes for all large enough odd integer $m$.
Indeed fortuitous sequences are known for 3, 15, 23, 29, 31, 33 and 37.
For small $n$ they can be found by trying all ${n-1\over2}!$ possibilities. (6h on a PC for \n=29).
For large $n$, search process must be improved.
Fortuitous sequence $S=ns_1ns_2ns_3$ yields fortuitous sequences 
$ns_2ns_3ns_1$, $ns_3ns_1ns_2$, $n\tilde s_3n\tilde s_2n\tilde s_1$, $n\tilde s_2n\tilde s_1n\tilde s_3$ and $n\tilde s_1n\tilde s_3n\tilde s_2$.
But these 6 sequences are not necessarily different.
There are only 3 different sequences if $s_1=\tilde s_1$, only 2 if $s_1=s_2$ and only 1 if $s_1=s_2=\tilde s_1$.
Indeed for \n=3 and \n=15 there is only one fortuitous sequence. Hence it verifies $s_1=s_2=\tilde s_1$.
So it makes sense to search only for small fortuitous sequences verifying $s_1=\tilde s_1$ or $s_1=s_2$,
and then combine them to get larger fortuitous sequences.
Note that symmetry is lost through combination.
Combining the triple palindromic sequence $S_{15}$ with itself, yields a sequence with no symmetry.

\titr{IIIb Palindromic fortuitous sequences}
When $s_1=\tilde s_1$, then $s_1$ is a palindrome containing exactly one occurrence of $n{-}1$. So $s_1=w~n{-}1~\tilde w$ is of odd length ${n-1\over2}$ and
$n\equiv 3\pmod 4$.
An exhaustive search of all $s_1=w~n{-}1~\tilde w$ tries ${n-1\over2}!/{n+1\over4}!$ cases
and finds readily all fortuitous sequences for \n=27 (0.16s), \n=31 (3.9s) and \n=35 (2m 2s).
E.g. it finds sequences of the form $(n~w_n~n{-}1~\tilde w_n)$ with\br
$w_{23}=($14  4  6 22 10  8 12 10 14 12 18 23 10 14 18  8 10)\br
$w_{27}=($22 16 10 18 26  6  4 12 14 22  6  4 10 27 12  4  8 22  4  6)\br
$w_{31}=($18  4 14 22  6  4 18 26  6  4 22 30  6  4 26 31 10  4 18  4 26  4 14)\br
$w_{35}=($ 6 22 18 10 32  8 26 22 34 22 26  8 32 10 18 22  6 35  6 22 18 10 32  8 26 22)

We may notice that if $f_{nwx\tilde w}=-I_n$ then $f_x=f_w^{-1}\circ-f_n\circ f_w$ is conjugate to
$-f_n$ and they have as many fixed points. If $n$ is odd, $-f_n$ has one fixed point ${n+1\over2}$. So $f_x$ has one fixed point and $x=n-1$.
If $n$ is even, $-f_n$ and $f_x$ have no fixed point, so $x=n$.
So the central flip of a palindromic flip sequence sorting $-f_n$ is always
either $f_{n-1}$ or $f_n$ depending on the parity of $n$, even if it is not a part of a fortuitous sequence.

\titr{IIIc Triple fortuitous sequences}
For $n\equiv 1 \pmod 4$ we cannot have $s_1=\tilde s_1$, but we may have $s_1=s_2=s_3$.
Then the stack of 2-clans is a signed permutation of order 3.
It is possible to search for all stacks of 2-clans of order 3, and keep among them those corresponding to fortuitous sequences.
Let $d(i)={\rm sgn}(i)(-1)^i$.
A stack of 2-clans of order 3 is a permutation of $\{-n,1{-}n,\ldots-1,0,1,2,\ldots n\}$ verifying the following conditions:
\halign{$#$\hss\quad&#\hss\cr
*\forall i,\; f(-i)=-f(i)                  &$f$ is a signed permutation.\cr
*f(n)=1                                    &The smallest pancake is right-side up at the bottom.\cr
*f(1)<0                                    &The top pancake is upside down.\cr
*\forall i,\; f(f(f(i)))=i                 &$f$ is a permutation of order 3.\cr
*\forall i,\; d(f(i))=d(i)                 &Pancake $i$ is oriented according to the parity of $i+\abs{f(i)}$.\cr
*\forall i,\; f(i-d(i))=f(i)+d(i)          &$f$ is made of clans of 2 pancakes.\cr
*\forall i\ne0,\; f(i+d(i))\ne f(i)-d(i)   &The clans of 2 pancakes do not form larger clans.\cr
}
Depth first search is effective : We pick up any $i$ such that $j=f(i)$ is known and $f(j)$ is unknown.
Then we shall try all possible values $k$ for $f(j)$.
The absolute value of $k$ must be taken in set $K=\{x{>}0~|~f(x){\rm~and~}f^{-1}(x){\rm~are~unknown}\}$.
The sign of $k$ is given by rule $d(k)=d(i)$, so $k=|k|d(|k|)d(i)$.
Then we force $f(j)=k$ and $f(k)=i$.
Furthermore whenever we force $f(x)=y$, we shall also force $f(-x)=-y$ and $f(x-d(x))=y+d(y)$
and if only one of the numbers $f(y)$ and $f^{-1}(x)$ is known, then both numbers are forced to the same value.
The first choice of $|k|$ is made in set $K=\{2,3,\dots,n-1\}$ since $f(n)=1$. After each choice of $k$
set $K$ loses elements $|k|$, $|k+1|$, $|k-1|$, $|j-d(j)|$ and $|i+d(i)|$.
Generally it loses at least 5 elements, since if for instance $|k+d(k)|$ was already excluded because $z=f(k+d(k))$ is known,
then we shall force $f(z)=j-d(j)$ and exclude other elements.
An exhaustive search of all stack of 2-clans of order 3, tries about $(n-2)(n-7)(n-12)(n-17)\ldots$ cases
and finds readily all triple fortuitous sequences for \n=29 (0.08s), \n=31 (0.23s), \n=33 (0.79s), \n=35 (2.8s) and \n=37 (10.5s). E.g.\br
(29 16  4 26  6 20  4 16 24  6 14 28 12  4 24$)^3$\hfill
(31 18  8 28  6 18  4 14  8  4 22  6 24 30  4 16$)^3$\break
(33 18  8 30  8  2 20  4 14  8  4 24  8 26 32  4 18$)^3$\hfill
(35 20  8  2 32  8 22  4 18  2 28  2  8 16 34 16  4 30$)^3$\break
(37 22 12  2 34  8 20  4 16  2 10  4 26  2  8 28 36  6 20$)^3$\br
Adding rule $f=f_n\circ f^{-1}\circ f_n$ gives readily all palindromic triple fortuitous sequences for \n=35, 39, 43 and 47(0.57s).
E.g. it gives $(n~w_n~n{-}1~\tilde w_n)^3$ with
$w_{35}=$(6  22 18 10 32  8 26 22),
$w_{39}=$(22 16 18 14 16 30 34 12 14),
$w_{43}=$(6  26 22 14 40 10 12 10 30 26) and
$w_{47}=$(6  28 24 16 44 12  2 14 10 32 28).

\titre{IV Even fortuitous sequences, patchworks of 2-clans and 2-blocks and their combination}
For even $n$ a sequence satisfying the lower bound is of length $3n/2+1$ and holds two $f_n$.
So we define an ``even fortuitous sequence'' as a flip sequence of length $3n/2+1$ of the form $S=ns_1ns_2$,
where $s_1$ and $s_2$ are sequences of positive integers lower than $n$, such that $f_S=-I_n$ and
stack $f=-f_{ns_1}$ obtained at the end of $s_1$, when applying $S$ from $-I_n$ or equivalently $s_1$ from $-f_n$,
is a ``patchwork'' of $n/2-1$ clans or blocks of 2 pancakes, placed on top of pancake 1 and below singleton $-n$.

So no pancake of this patchwork has potential 3 or -3, and
every flip of $s_1$ increases potential by 4, and breaks an anti-adjacency.
As well every flip of $s_2$ creates an adjacency.

We may notice that $-f^{-1}$ and $f_n\circ f^{-1}\circ f_n$ and $-f_n\circ f\circ f_n$ are the patchworks of
fortuitous sequences $ns_2ns_1$ and $n\tilde s_1n\tilde s_2$ and $n\tilde s_2n\tilde s_1$ respectively.
Hence $\tilde s_1$, $\tilde s_2$, $s_1$, and $s_2$ are
the only flip sequences adding an adjacency at each step when starting respectively from stacks
$-f$, $f^{-1}$, $-f_n\circ f^{-1}$ and $f_n\circ f$.
So patchwork $f$ characterizes even fortuitous flip sequence $S$.

We may roughly count the even fortuitous sequences.
The number of patchworks is ${n-2\over 2}!2^{n-2}$, since there are
${n-2\over 2}!$ possible orders for the ${n-2\over 2}$ pieces of 2 pancakes,
which have each 4 possible orientations (E.g.  [2,3], [-2,-3], [3,2] or [-3,-2]).
About a quarter of these patchworks are not suitable since they have too large blocks or clans.
When choosing a patchwork, $s_1$ and $s_2$ are imposed.
Indeed there is only one way to create an adjacency by a flip.
But this flip has about one chance out of two to exist, whether the pancake matching the first one shows its right side or not.
Hence the probability that $s_2$ exists is about $2^{-\abs{s_2}}$, where $\abs{s_2}$, the length of $s_2$, is the
number of missing adjacencies in the patchwork.
As well the probability that $s_1$ exists is about $2^{-\abs{s_1}}$, where $\abs{s_1}$ is the
number of missing anti-adjacencies in the patchwork.
Obviously $\abs{s_1}+\abs{s_2}={3n\over 2}-1$,
so the expected number of fortuitous sequences may be estimated as
${3\over 4}{n-2\over2}!2^{-1-n/2}\to\infty$.
This suggests, that fortuitous sequences exist for every large enough even integer $n$.
This can be proved in two different ways, that we shall both use.
Either we look for regular patterns in patchworks of small even fortuitous sequences and
design according to these patterns a patchwork yielding a fortuitous sequence for any large even $n$.
Or we exhibit small even fortuitous sequences with symmetries, like palindromic or double fortuitous sequences
and use method of Heidari et al. to combine two even fortuitous sequences for $n_1$ and $n_2$ pancakes
into a fortuitous sequence for $n_1+n_2-2$ pancakes.
E.g. even fortuitous sequences\br
(26 20 14 16 11  3 24 11 16  8 19  7 13 11 25  8 21 18  3 15 26 15  3 18  5 21  8 25 13  5 19  8 16 11 24  3 11 14 10 18)\br
(28 22  2 16 18 13  3 26 11 18  2 10 21  7 15 13 27  8 23 20  3 17 28 15  3 18  5 21  8 27 15  5 21 10  2 18 11 26  3 13 16 10 20)
yield patchworks:\br
[-26  6  7 -21 -20 -23 -22 -16 -17 -3 -2 -13 -12 15 14 25 24 10 11  5  4  9  8 -18 -19$|$1]\br
[-28$|$6  7 -9 -8 -23 -22 -25 -24 -18 -19 -3 -2 -15 -14 17 16 27 26 12 13  5  4 11 10 -20 -21  1]\br
In the second stack, the top singleton, -28(=-$n_2$), is replaced by the first stack without its final 1.
Also in the first stack, pancakes 2$\ldots$ 26(=$n_1$) become 28(=$n_2$)$\ldots$ 52(=$n_1+n_2-2$)
by increasing their numbers by 26(=$n_2-2$).
In the second stack, the pancakes keep their numbers 1$\ldots$ 27(=$n_2-1$).
This yields patchwork\br
[-52 32 33-47-46-49-48-42-43-29-28-39-38 41 40 51 50 36 37 31 30 35 34-44-45$|$6  7 -9 -8-23-22-25-24-18-19 -3 -2-15-14 17 16 27 26 12 13  5  4 11 10-20-21  1]\br
and even fortuitous sequence\br
(52 20 14 16 11  3 24 11 16  8 19  7 13 11 46  2 16 18 13  3 50 35 42  2 10 45  7 39 37 51  8 47 44  3 41  8 21 18  3 15 52 15  3 18  5 21  8 41  3 44 31 47  8 51 39  5 45  8 42 37 50  3 11 14 10 44 15  5 21 10  2 18 11 26  3 13 16 10 20)

Since there is a fortuitous sequence for 26 pancakes, any fortuitous sequence for $n$ pancakes yields fortuitous sequences
for $n+24$, $n+48$, $n+72$, etc..
Moreover, the even fortuitous sequences for 26 and 28 pancakes yield even fortuitous sequences for all \n=24$i$+26$j$+2,
i.e. for all large enough even integer $n$.
Indeed fortuitous sequences are known for 2 and all even integers from 26 to 48.
This yield fortuitous sequences for all even $n\ge 26$.
\vfill\eject
\titr{IVa Searching for even fortuitous sequences}
Patchworks are permutations of $\{-n,1{-}n,\ldots-1,0,1,2,\ldots n\}$ verifying the following conditions:
\halign{$#$\hss\quad&#\hss\cr
*\forall i,~f(-i)=-f(i)            &$f$ is a signed permutation.\cr
*f(n)=1                            &The smallest pancake is right-side up at the bottom.\cr
*f(1)=-n                           &The largest pancake is upside down on the top.\cr
*\forall i,~f(i+d(i))=f(i)+d(f(i)) &$f$ is made of clans and blocks of 2 pancakes.\cr
*\forall i\ne0,~2f(i)\ne f(i-1)+f(i+1) &Clans and blocks of 2 pancakes do not form larger clans or blocks.\cr
}
E.g. $f(4)=-11$ implies $f(-4)=11$, $f(5)=-10$, $f(-5)=10$, $f(6)\ne-9$ and $f(3)\ne-12$.
We perform a depth first search to find all even fortuitous sequences.
At the start patchwork $f$ and stacks $-f$, $f^{-1}$, $-f_n\circ f^{-1}$ and $f_n\circ f$ have most of their values unknown.
To make sure a flip on one of the four stacks creates an adjacency, we may have to define some values of $f$.
We choose the stack which minimizes the number of unknown values of $f$ to define.
E.g. a stack of the form $[x\ldots y\ldots]$ is the best choice, if values $x$ and $y=1-x$ are known,
while a stack with an unknown top pancake is the worst choice.
Once the stack is chosen we try any possible flip on it and choose again a stack.
When the top of a stack is 1 or $-n$ we may force the stack to be $I_n$ or $f_n$.
This depth first search finds even fortuitous sequences for $n=26$(in 1m 23s) and $n=28$(in 31m).
Indeed it finds all solutions in 20m and 6h respectively, but for 26 as well as for 28,
there are only 4 solutions, which yield one another :
$ns_1ns_2$, $ns_2ns_1$, $n\tilde s_1n\tilde s_2$ and $n\tilde s_2n\tilde s_1$.
We may look for similarities between a solution for 26 and a solution for 28.
So we force $f(2)=11$, $f(-4)=21$, $f(-6)=23$, $f(-8)=2$, $f(-10)=18$, $f(1-n)=16$, $f(n-3)=6$ and $f(n-5)=n-1$.
This gives readily solutions for 30, 32, 34 and 38.
Then looking for other similarities between those solutions, we impose\br
$f=[-n$ 11 10 -21 -20 -23 -22 -2 -3 -18 -19 $w_n$ 13 12 -15 -14  8  9  5  4 $w'_n$ $n{-}2$ $n{-}1$ 7  6 -17 -16  1]\br
where $w_n$ is an initial segment of length multiple of 4, of infinite sequence :
-26 -27 -24 -25$|$-30 -31 -28 -29$|$-34 -35 -32 -33$|$-38 -39 -36 -37$|$-42 -43 -40 -41$|$-46 -47 -44 -45$|$-50 -51 -48 -49$|$-54 -55 -52 -53 $\ldots$
and $w'_n$ is empty if 4 divides $n$ and $w'_n=(3{-}n,4{-}n)$ otherwise.
The first patchworks and flip sequences are :
\def\bbb{~~~~~~~~} \def\bbbb{\bbb\bbb} 

\noindent
[-26 11 10 -21 -20 -23 -22 -2 -3 -18 -19 \bbbb\bbbb              13 12-15-14  8  9  5  4 \bbb  24 25  7  6 -17 -16  1]\br
[-28 11 10 -21 -20 -23 -22 -2 -3 -18 -19 \bbbb\bbbb              13 12-15-14  8  9  5  4-25-24 26 27  7  6 -17 -16  1]\br
[-30 11 10 -21 -20 -23 -22 -2 -3 -18 -19-26-27-24-25 \bbbb       13 12-15-14  8  9  5  4 \bbb  28 29  7  6 -17 -16  1]\br
[-32 11 10 -21 -20 -23 -22 -2 -3 -18 -19-26-27-24-25 \bbbb       13 12-15-14  8  9  5  4-29-28 30 31  7  6 -17 -16  1]\br
[-34 11 10 -21 -20 -23 -22 -2 -3 -18 -19-26-27-24-25-30-31-28-29 13 12-15-14  8  9  5  4 \bbb  32 33  7  6 -17 -16  1]\br
[-36 11 10 -21 -20 -23 -22 -2 -3 -18 -19-26-27-24-25-30-31-28-29 13 12-15-14  8  9  5  4-33-32 34 35  7  6 -17 -16  1]
\halign{#&~#$\ldots$\hss\cr
(26 18 10 14 11 3 24 11 16   \hss                           8 19 5&13 25 ~8 21 5 18\cr
(28 20 10 16 13 3 26 11 18 2 \hss                          10 21 5&15 27 ~8 21 5 18\cr
(30 22 14 18 15 3 28 11 16   \hss                    21 23 10 21 5&15 29 10 23 5 20 26 24\cr
(32 24 14 20 17 3 30 11 18 2 \hss                    23 25 12 23 5&17 31 10 23 5 20 26 24\cr
(34 26 18 22 19 3 32 11 16   \hss              25 27 23 25 12 23 5&17 33 12 25 5 22 28 26 30 28\cr
(36 28 18 24 21 3 34 11 18 2 \hss              27 29 25 27 14 25 5&19 35 12 25 5 22 28 26 30 28\cr
(40 32 22 28 25 3 38 11 18 2 \hss        31 33 29 31 27 29 16 27 5&21 39 14 27 5 24 30 28 32 30 34 32\cr
(44 36 26 32 29 3 42 11 18 2 \hss  35 37 33 35 31 33 29 31 18 29 5&23 43 16 29 5 26 32 30 34 32 36 34 38 36\cr
(48 40 30 36 33 3 46 11 18 2 39 41 37 39 35 37 33 35 31 33 20 31 5&25 47 18 31 5 28 34 32 36 34 38 36 40 38 42 40\cr}
\halign{$\ldots$#&&~#\cr
3 15 26 15  3 18 21  8 25 11 13  7 19 ~8\hss16 11 24 &3 11            \hss16&14 \hss               20)\cr
3 15 28 17  3 20 23  8 27 13 15  7 21 10  2 18 11 26 &3 13            \hss18&16 \hss             2 22)\cr
3 15 30 19  3 22 25  8 29 15 17  7 23 12\hss20 11 28 &3 11 15         \hss20&18 \hss         11  ~ 24)\cr
3 15 32 21  3 24 27  8 31 17 19  7 25 14  2 22 11 30 &3 13 17         \hss22&20 \hss         13  2 26)\cr
3 15 34 23  3 26 29  8 33 19 21  7 27 16\hss24 11 32 &3 11 15 19      \hss24&22 \hss      15 11  ~ 28)\cr
3 15 36 25  3 28 31  8 35 21 23  7 29 18  2 26 11 34 &3 13 17 21      \hss26&24 \hss      17 13  2 30)\cr
3 15 40 29  3 32 35  8 39 25 27  7 33 22  2 30 11 38 &3 13 17 21 25   \hss30&28 \hss   21 17 13  2 34)\cr
3 15 44 33  3 36 39  8 43 29 31  7 37 26  2 34 11 42 &3 13 17 21 25 29\hss34&32 \hss25 21 17 13  2 38)\cr
3 15 48 37  3 40 43  8 47 33 35  7 41 30  2 38 11 46 &3 13 17 21 25 29 33 38&36  29 25 21 17 13  2 42)\cr}
This provides even fortuitous sequences for all $n\ge26$.
\vfill\eject
\titr{IVb Double fortuitous sequences}
For odd $n$ a fortuitous sequence holds 3 occurrences of $f_n$ and may be of the form $(ns)^3$.
As well, an even fortuitous sequence holds 2 occurrences of $f_n$ and may be of the form $(ns)^2$.
Then its length $3n/2+1$ is even and $n\equiv 2\pmod4$.

An even fortuitous sequence is double whenever its patchwork verifies $f^{-1}=-f$.
With this extra rule depth first search for even fortuitous sequences finds all
double fortuitous sequences for $n=30$(in 0.75s), $n=34$(in 16.6s) and $n=38$(in 9m).
Some patchworks for 30 have similarities with some patchworks for 34, such as $f(2)=8$.
When adding this condition, computation is much faster.
So we look for other similarities between patchworks for 30, 34 and 38 and add these conditions. Etc.
At last we impose $f(2)=8$, $f(n/2+8)=4$, $f(6)=n/2$, $f(n/2+1)=n/2+3$, $f(n/2+5)=n-1$
and $f(n/2+11+i)=14+i$ and $f(n/2+13+i)=12+i$ for $i=0$, 4, 8, $\ldots$ $i<n/2-21$.
So we get double fortuitous sequences for $n=30$, 34, 38, 42, $\ldots$
E.g. patchworks :
\halign{[#$\ldots$\cr
-30  8  9-23-22 15 14 -2 -3      \hss                                -26-27-24-25 -7\cr
-34  8  9-25-24 17 16 -2 -3      \hss        -28-29     \hss          31 30-26-27 -7\cr
-38  8  9-27-26 19 18 -2 -3-34-35\hss                          -30-31-28-29-32-33 -7\cr
-42  8  9-29-28 21 20 -2 -3-38-39\hss                    -32-33 35 34-30-31-36-37 -7\cr
-46  8  9-31-30 23 22 -2 -3-42-43-36-37-34-35  \hss            -38-39-32-33-40-41 -7\cr
-50  8  9-33-32 25 24 -2 -3-46-47-38-39-36-37  \hss      -40-41 43 42-34-35-44-45 -7\cr
-54  8  9-35-34 27 26 -2 -3-50-51-40-41-38-39-44-45-42-43 \hss -46-47-36-37-48-49 -7\cr
-58  8  9-37-36 29 28 -2 -3-54-55-42-43-40-41-46-47-44-45-48-49 51 50-38-39-52-53 -7\cr
}\halign{$\ldots$#]\cr
-6 18 19-16-17 29 28  5  4     \hss                                  12 13 10 11-21-20  1\cr
-6 20 21-18-19 33 32  5  4   \hss   14 15 10 11-13-12     \hss                  -23-22  1\cr
-6 22 23-20-21 37 36  5  4 14 15   \hss                        12 13 16 17 10 11-25-24  1\cr
-6 24 25-22-23 41 40  5  4 16 17   \hss                  12 13-15-14 18 19 10 11-27-26  1\cr
-6 26 27-24-25 45 44  5  4 18 19 14 15 12 13   \hss            16 17 20 21 10 11-29-28  1\cr
-6 28 29-26-27 49 48  5  4 20 21 14 15 12 13   \hss      16 17-19-18 22 23 10 11-31-30  1\cr
-6 30 31-28-29 53 52  5  4 22 23 14 15 12 13 18 19 16 17 \hss  20 21 24 25 10 11-33-32  1\cr
-6 32 33-30-31 57 56  5  4 24 25 14 15 12 13 18 19 16 17 20 21-23-22 26 27 10 11-35-34  1\cr
}
give double fortuitous sequences\br
$($30 22 14  7  3 24  7 27 13 15 10 29 15 20 18  5 23 14 20 13 28  3 19$)^2$\br
$($34 26 16  7  3 28  9 31 13 17  2 12 33 17 24  2 22  5 27 18  2 24 15 32  3 21$)^2$\br
$($38 30 18  7  3 32  7 35 15 17 15 19 14 37 19 26 28 26 22  9 31 18 26 28 26 17 36  3 23$)^2$\br
$($42 34 20  7  3 36  7 39 15 19  2 17 21 16 41 21 28 32  2 30 24 11 35 20 28 32  2 30 19 40  3 25$)^2$\br
$($46 38 22  7  3 40  7 43 15 19 21 19 15 23 18 45 23 34 36 32 34 32 26 13 39 22 32 34 32 36 34 21 44  3 27$)^2$\br
$($50 42 24  7  3 44  7 47 15 21 23 21  2 17 25 20 49 25 36 40  2 36 38 36 28 15 43 24 34 36 34 40  2 38 23 48  3 29$)^2$\br
$($54 46 26  7  3 48  7 51 15 23 25 21 23 21 15 27 22 53 27 42 44 38 40 38 42 40 30 17 47 26 36 40 42 40 36 44 42 25 52  3 31$)^2$\br
$($58 50 28  7  3 52  7 55 15 25 27 23 25 23  2 17 29 24 57 29 44 48  2 42 44 42 46 44 32 19 51 28 38 42 44 42 38 48  2 46 27 56  3 33$)^2$
\titr{IVc Palindromic even fortuitous sequences centered on $n$}
An even fortuitous sequence is of the form $ns_1ns_2=n\tilde s_2n\tilde s_1$ whenever $f=-f_n\circ f\circ f_n$.
We add this rule as well as $f(2)=n/2-2$, $f(6)=-2$, $f(n/2+4)=n/2+5$.
Furthermore when $n\ge34$ we add rules $f(i)=-i$ and $f(i+2)=2-i$ for $i=n/2-7$ and $i=4$, 8, 12, $\dots8\ent{n/16}-12$.
Then depth first search for even fortuitous sequences finds readily
even palindromic fortuitous sequences of the form $nw_nn\tilde w_n$ for any $n\ge30$, if $n\equiv2\pmod4$. E.g.~:\br
$w_{30}$=(24 14 12 20 15  3 28 11 16 14 20 14 23  5 17 29  8 23  7 20  3 17)\br
$w_{34}$=(28 30 10 24 26 23  3 32  7 10  8 14 33 23 10  8 14 17  7 31 16 25 22  3 19)\br
$w_{38}$=(32 34 12  2 26 28 25  3 36  9 12 10 18 37 25 12 10  2 16 19  9 35  2 18 27 24  3 21)\br
$w_{42}$=(36 38 14 34 30 24 26 23  3 40 15 18 16 22 41 31 18 16 22 25 15 39 33 35 31 33 20 29 26  3 23)\br
$w_{46}$=(40 42 16 36 32  2 26 28 25  3 44 17 20 18 26 45 33 20 18  2 24 27 17 43  2 37 39 35 37 22 31 28  3 25)\br
$w_{50}$=(44 46 42 44 40 42 22 36 38 35  3 48 11 14 12 18 22 26 49 35 18 16 20 18 22 20 26 29 11 47 43 39 16 33 30  3 27)\br
$w_{54}$=(48 50 46 48 44 46 24  2 38 40 37  3 52 13 16 14 22 26 30 53 37 20 18 22 20 24 22  2 28 31 13 51 47 43  2 18 35 32  3 29)\br
$w_{58}$=(52 54 50 52 48 50 26 46 42 36 38 35  3 56 19 22 20 26 30 34 57 43 26 24 28 26 30 28 34 37 19 55 51 47 41 43 39 41 20 37 34  3 31)\br
$w_{62}$=(56 58 54 56 52 54 28 48 44  2 38 40 37  3 60 21 24 22 30 34 38 61 45 28 26 30 28 32 30  2 36 39 21 59 55 51  2 45 47 43 45 22 39 36  3 33)\br
$w_{66}$=(60 62 58 60 56 58 54 56 52 54 34 48 50 47  3 64 15 18 16 22 26 30 34 38 65 47 26 24 28 26 30 28 32 30 34 32 38 41 15 63 59 55 51 47 16 41 38  3 35)\br
[-30 13 12-23-22 -2 -3 -6 -7-27-26-11-10 16 17-14-15 21 20  5  4 24 25 28 29  9  8-19-18  1]
\halign{[#&\hss#$\ldots$\cr
-34 15 14 -4 -5 -2 -3     \hss                                        28 29-10-11&-8~~-9-13-12 18 19-16\cr
-38 17 16 -4 -5 -2 -3     \hss                        -31-30     \hss 32 33-12-13&-10-11-15-14 20 21-18\cr
-42 19 18 -4 -5 -2 -3     \hss                      32 33 -8 -9  \hss 36 37-14-15&-12-13-17-16 22 23-20\cr
-46 21 20 -4 -5 -2 -3     \hss                  -35-34 36 37 -8 -9\hss40 41-16-17&-14-15-19-18 24 25-22\cr
-50 23 22 -4 -5 -2 -3 -8 -9 -6 -7-12-13-10-11 \hss              \hss  36 37-18-19&-16-17-21-20 26 27-24\cr
-54 25 24 -4 -5 -2 -3 -8 -9 -6 -7-12-13-10-11 \hss    -39-38    \hss  40 41-20-21&-18-19-23-22 28 29-26\cr
-58 27 26 -4 -5 -2 -3 -8 -9 -6 -7-12-13-10-11 \hss  40 41-16-17 \hss  44 45-22-23&-20-21-25-24 30 31-28\cr
-62 29 28 -4 -5 -2 -3 -8 -9 -6 -7-12-13-10-11\hss43-42 44 45-16-17\hss48 49-24-25&-22-23-27-26 32 33-30\cr
-66 31 30 -4 -5 -2 -3 -8 -9 -6 -7-12-13-10-11-16-17-14-15-20-21-18-19 44 45-26-27&-24-25-29-28 34 35-32\cr
}\halign{$\ldots$#]\cr
-17 23 22 26 27 24 25 -6 ~-7                         \hss                    32 33 30 31-21-20  1\cr
-19 25 24 28 29 26 27 -6 ~-7 \hss      9  8          \hss                    36 37 34 35-23-22  1\cr
-21 27 26 30 31 28 29 -6 ~-7 \hss  34 35-10-11       \hss                    40 41 38 39-25-24  1\cr
-23 29 28 32 33 30 31 -6 ~-7\hss38 39-10-11 13 12    \hss                    44 45 42 43-27-26  1\cr
-25 31 30 34 35 32 33-14-15                  \hss   40 41 38 39 44 45 42 43 48 49 46 47-29-28  1\cr
-27 33 32 36 37 34 35-14-15 \hss     17 16   \hss   44 45 42 43 48 49 46 47 52 53 50 51-31-30  1\cr
-29 35 34 38 39 36 37-14-15 \hss  42 43-18-19 \hss  48 49 46 47 52 53 50 51 56 57 54 55-33-32  1\cr
-31 37 36 40 41 38 39-14-15\hss46 47-18-19 21 20\hss52 53 50 51 56 57 54 55 60 61 58 59-35-34  1\cr
-33 39 38 42 43 40 41-22-23 48 49 46 47 52 53 50 51 56 57 54 55 60 61 58 59 64 65 62 63-37-36  1\cr
}
\titr{IVd Palindromic even fortuitous sequences centered on $n-1$}
An even fortuitous sequence is of the form $ns_1ns_2=n\tilde s_1n\tilde s_2$ whenever $f=f_n\circ f^{-1}\circ f_n$.
When adding this rule as well as $f(6)=11-n$, $f(8)=-3$, $f(15)=-4$ and $f(i)=i+1-n$ and $f(i+2)=i-1-n$ for $i=16$, 20, 24, $\ldots 8\ent{n/8}-16$,  
depth first search for even fortuitous sequences finds readily even palindromic fortuitous sequences for any $n\ge34$, if $n\equiv2\pmod4$.
After rotation these sequences take the form $(n{-}1~w_n~n{-}1~\tilde w_n)$. E.g.~:\br
$w_{34}$=(24 12 20 27  9 23 32 26 30 17 25 34 27 17  6  8 30  9 20 32  6 27 12 19 11)\br
$w_{38}$=(26 12 22 31 11 27 19 17 36 30 34 17 29 38 27 17  6  8 34 13 24 36  6 31 16  7 23 11)\br
$w_{42}$=(28 12 24 35 31 27  9 23 40 34 38 25 33 42 35 25 14 16 12 14 10 12 38  9 20 40 10 35 12 23 15)\br
$w_{46}$=(30 12 26 39 35 31 11 27 19 17 44 38 42 25 37 46 35 25 14 16 12 14 10 12 42 13 24 44 10 39 16  7 27 15)\br
$w_{50}$=(32 12 28 43 39 35 31 27  9 23 48 42 46 33 41 50 43 33 22 24 20 22 18 20 16 18 14 16 46  9 20 48 14 43 12 27 19)\br
$w_{54}$=(34 12 30 47 43 39 35 31 11 27 19 17 52 46 50 33 45 54 43 33 22 24 20 22 18 20 16 18 14 16 50 13 24 52 14 47 16  7 31 19)
\halign{[#$\dots$\cr
-34  7  6 11 10-23-22 -3 -2 14 15-29-28 -5 -4  \hss                                          -17-16-19-18\cr
-42  7  6 11 10-31-30 -3 -2 14 15-37-36 -5 -4  \hss                  -25-24-27-26-21-20-23-22-17-16-19-18\cr
-50  7  6 11 10-39-38 -3 -2 14 15-45-44 -5 -4-33-32-35-34-29-28-31-30-25-24-27-26-21-20-23-22-17-16-19-18\cr
}\halign{$\dots$#]\cr
 24 25 -8 -9 31 30-12-13 33 32-21-20-27-26  1\cr
 32 33 -8 -9 39 38-12-13 41 40-29-28-35-34  1\cr
 40 41 -8 -9 47 46-12-13 49 48-37-36-43-42  1\cr
}\halign{[#$\dots$\cr
-38 11 10 15 14-27-26 -3 -2 18 19-33-32 -5 -4  \hss                                          -21-20-23-22\cr
-46 11 10 15 14-35-34 -3 -2 18 19-41-40 -5 -4  \hss                  -29-28-31-30-25-24-27-26-21-20-23-22\cr
-54 11 10 15 14-43-42 -3 -2 18 19-49-48 -5 -4-37-36-39-38-33-32-35-34-29-28-31-30-25-24-27-26-21-20-23-22\cr
}\halign{$\dots$#]\cr
 28 29-12-13 35 34-16-17 37 36  7  6  9  8-25-24-31-30  1\cr
 36 37-12-13 43 42-16-17 45 44  7  6  9  8-33-32-39-38  1\cr
 44 45-12-13 51 50-16-17 53 52  7  6  9  8-41-40-47-46  1\cr
}

\titr{IVe Palindromic double fortuitous sequences}
An even fortuitous sequence is of the form $(ns)^2=(n\tilde s)^2$ whenever $-f=f^{-1}=f_n\circ f\circ f_n$.
When adding this rule
as well as $f(n-1)=1-8\ent{n/16}$ and $f(i)=i+1-n$ and $f(i+2)=i-1-n$ for $i=4$, 8, 12, $\ldots 4\ent{n/8}-20$
depth first search for even fortuitous sequences finds readily double palindromic fortuitous sequences
of the form $(n~w_n~n{-}1~\tilde w_n)^2$ for any $n\ge34$, if $n\equiv2\pmod8$. E.g.~:\br
$w_{34}$=(14  5  3 30 14 10  5  3 18  6 21 10)\br
$w_{42}$=(22 16 18 28 19 17 22 12 32 27 37 17 23 31 27)\br
$w_{50}$=(22 16 18 14 16  9  3 38 14 10 41 45  9  3 22  6 29 18)\br
$w_{58}$=(30 24 26 22 24 20 22 36 23 21 26 12 40 35 45 49 53 25 31 43 39)\br
$w_{66}$=(30 24 26 22 24 20 22 18 20 13  3 46 14 10 49 53 57 61 13  3 26  6 37 26)\br
$w_{74}$=(38 32 34 30 32 28 30 26 28 24 26 44 27 25 30 12 48 43 53 57 61 65 69 33 39 55 51)
\halign{[#&\hss#&\hss#$\dots$\cr
-34 20 21 \hss                                           -26-27 17 16-30-31-12-13 10 11 32 33& -7& -6 29 28\cr
-50 28 29 \hss                   -45-44-47-46-41-40-43-42-34-35 25 24-38-39-20-21 18 19 48 49&-15&-14 37 36\cr
-66 36 37-61-60-63-62-57-56-59-58-53-52-55-54-49-48-51-50-42-43 33 32-46-47-28-29 26 27 64 65&-23&-22 45 44\cr
}\halign{$\dots$#]\cr
 -2 -3-24-25 22 23 ~4 ~~5-19-18 ~8 ~~9              \hss                              -14-15  1\cr
 -2 -3-32-33 30 31 12  13-27-26 16  17              \hss        9  8 11 10  5  4  7  6-22-23  1\cr
 -2 -3-40-41 38 39 20  21-35-34 24  25 17 16 19 18 13 12 15 14  9  8 11 10  5  4  7  6-30-31  1\cr
}\halign{[#&\hss#&\hss#$\dots$\cr
-42 20 21 \hss                                           -37-36-39-38 14 15 27 26 19 18& -8& -9 33 32-13-12\cr
-58 28 29 \hss                   -53-52-55-54-49-48-51-50-45-44-47-46 22 23 35 34 27 26&-16&-17 41 40-21-20\cr
-74 36 37-69-68-71-70-65-64-67-66-61-60-63-62-57-56-59-58-53-52-55-54 30 31 43 42 35 34&-24&-25 49 48-29-28\cr
}\halign{$\dots$#]\cr
 -2 -3 40 41 31 30-11-10 34 35-25-24-17-16-28-29  \hss                                            5  4  7  6-22-23  1\cr
 -2 -3 56 57 39 38-19-18 42 43-33-32-25-24-36-37  \hss                   13 12 15 14  9  8 11 10  5  4  7  6-30-31  1\cr
 -2 -3 72 73 47 46-27-26 50 51-41-40-33-32-44-45 21 20 23 22 17 16 19 18 13 12 15 14  9  8 11 10  5  4  7  6-38-39  1\cr
}

\titre{V Generalized odd fortuitous sequences with patchworks}
For odd $n$, fortuitous sequences may be generalized by replacing the stack of 2-clans placed on top of singleton 1,
by a patchwork of 2-clans and 2-blocks (placed on top of singleton 1).
So we define a ``generalized odd fortuitous sequence'' as a flip sequence of length ${3n+3\over2}$ of the form
$S=ns_1ns_2ns_3$ where $s_1$, $s_2$ and $s_3$ are
sequences of positive integers lower than $n$, such that $f_S=-I_n$ and
stack $f=-f_{ns_1}$ obtained at the end of $s_1$, when applying $S$ from $-I_n$ or equivalently $s_1$ from $-f_n$,
is a ``patchwork'' of ${n-1\over2}$ clans or blocks of 2 pancakes, placed on top of pancake 1.

So no pancake of this patchwork has potential 3 or -3, and
every flip of $s_1$ increases potential by 4, and breaks an anti-adjacency.
As well every flip of $s_2$ and of $s_3$ creates an adjacency.
We may notice that $h=f_n\circ f^{-1}\circ f_n$ is the patchwork of
generalized fortuitous sequence $n\tilde s_1n\tilde s_3n\tilde s_2$,
and $\tilde s_1$, $\tilde s_3$, $s_1$, and $s_2$ are
the only flip sequences of their lengths adding an adjacency at each step when starting respectively from stacks
$-f$, $f^{-1}$, $-f_n\circ f^{-1}$ and $f_n\circ f$.
So patchwork $f$ characterizes $s_1$, since its length is the number of missing anti-adjacencies in $f$.

Sequence $s_1$ derives from $f$ as $\tilde s_1$ derives from $h$.
Hence $f=h$ if and only if $s_1=\tilde s_1$, i.e. if $s_1$ is a palindrome.
E.g. for $n=15$ or $n=19$ there is only one generalized fortuitous sequence.
Hence $f=h$ and $s_1$ is a palindrome and $s_3=\tilde s_2$.  Hence $s_3ns_1ns_2$ is a palindrome.

Any fortuitous sequence is a generalized fortuitous sequence. Hence there are more generalized fortuitous sequences than fortuitous sequences and
hence more chance to find generalized fortuitous sequences than fortuitous sequences.
E.g. for $n=19$, there is no fortuitous sequence, but there is a generalized fortuitous sequence.
As well for $n=25$, no fortuitous sequence is known, but a generalized fortuitous sequence is easily found.
The patchworks of the generalized fortuitous sequences are the permutations of $\{-n,1{-}n,\ldots-1,0,1,2,\ldots n\}$
verifying the following conditions:
\halign{$#$\hss\quad&#\hss\cr
*\forall i,~f(-i)=-f(i)            &$f$ is a signed permutation.\cr
*f(n)=1                            &The smallest pancake is right-side up, at the bottom.\cr
*\forall i,~f(i-d(i))=f(i)+d(f(i)) &$f$ is made of clans and blocks of 2 pancakes.\cr
*\forall i\ne0,~2f(i)\ne f(i-1)+f(i+1) &Clans and blocks of 2 pancakes do not form larger clans or blocks.\cr
}
We perform a depth first search to find all generalized fortuitous sequences.
At the start, patchwork $f$ and stacks $-f$, $f^{-1}$, $-f_n\circ f^{-1}$ and $f_n\circ f$ have most of their values unknown.
To make sure a flip on one of the four stacks creates an adjacency, we may have to define some values of $f$.
We choose the stack which minimizes the number of unknown values of $f$ to define.
E.g. a stack of the form $[x\ldots y\ldots]$ is the best choice, if values $x$ and $y=1-x$ are known,
while a stack with an unknown top pancake is the worst choice.
Once the stack is chosen, we try any possible flip on it and choose again a stack.
When the top of a stack is 1 or $-n$ we may force the stack to be $I_n$ or $f_n$, as we do for even $n$.
So we use the same procedure for all $n$.
Note however that for two of the stacks, it shortens the path to the only solution when it exists,
but for the two other stacks it is only a tricky way to cut off useless computations.

This depth first search finds all the generalized fortuitous sequences for $n=15$, 19 and 25(in 6m 6s).
We may look for similarities between a solution for 15 and a solution for 19 and impose them.
This gives readily solutions for 15, 19, 23 and 27.
Then looking for other similarities between those solutions, we impose, when $n\equiv3\pmod4$ and $n\ge 15$, that
$f=[$-8 -9 $w_n$ -12 -13 -2 -3 1-{\n} -{\n} 5 4 7 6 -10 -11 1],
where $w_n$ is a terminal segment of length $n-15$, which is multiple of 4, of infinite sequence :\br
$\ldots$ -39 -38 -41 -40$|$-35 -34 -37 -36$|$-31 -30 -33 -32$|$-27 -26 -29 -28$|$-23 -22 -25 -24$|$-19 -18 -21 -20$|$-15 -14 -17 -16\br
These fortuitous sequences are of the form $(n~u_n~n{-}1~\tilde u_n)$ since $f=h$. The first patchworks and $u_n$ are :\br
\def\bbb{~~~~~~~~~~~~~~~~~~~~}
[-8 -9 \bbb\bbb\bbb\bbb                                                -12 -13 -2 -3 -14 -15 5 4 7 6 -10 -11 1]\br
[-8 -9 \bbb\bbb\bbb                                    -15 -14 -17 -16 -12 -13 -2 -3 -18 -19 5 4 7 6 -10 -11 1]\br
[-8 -9 \bbb\bbb                        -19 -18 -21 -20 -15 -14 -17 -16 -12 -13 -2 -3 -22 -23 5 4 7 6 -10 -11 1]\br
[-8 -9 \bbb            -23 -22 -25 -24 -19 -18 -21 -20 -15 -14 -17 -16 -12 -13 -2 -3 -26 -27 5 4 7 6 -10 -11 1]\br
[-8 -9 -27 -26 -29 -28 -23 -22 -25 -24 -19 -18 -21 -20 -15 -14 -17 -16 -12 -13 -2 -3 -30 -31 5 4 7 6 -10 -11 1]
\halign{#&&\hss~#\cr
$u_{15}=($10&               4& 6 14  6  4 10 15 10  4&                              &  6$)$\cr
$u_{19}=($14&            7  4&10 18  6  4 10 19 14  4&                         9  11&  8$)$\cr
$u_{23}=($18&        11  7  4&14 22  6  4 10 23 18  4&                13  15  11  13& 10$)$\cr
$u_{27}=($22&    15  11  7  4&18 26  6  4 10 27 22  4&        17  19  15  17  13  15& 12$)$\cr
$u_{31}=($26&19  15  11  7  4&22 30  6  4 10 31 26  4&21  23  19  21  17  19  15  17& 14$)$\cr}

As well, when $n\equiv1\pmod4$ and $n\ge 25$, we impose\br
$f=[$-8 -9 \n-3 \n-2 \n-7 \n-6 -12 -13 -2 -3 $w_n$ 5  4 {1-\n} {-\n} $w'_n$ {5-\n} {4-\n} 7 6 -10 -11  1],
where $w_n$ is a terminal segment of even length ${n-21\over 2}$ of infinite sequence:
$\ldots$ 49 48$|$45 44$|$41 40$|$37 36$|$33 32$|$29 28$|$25 24$|$21 20$|$17 16\br
and $w'_n$ is an initial segment of even length ${n-21\over 2}$ of infinite sequence:
15 14$|$19 18$|$23 22$|$27 26$|$31 30$|$35 34 $\ldots$\br
E.g. the patchworks and flip sequences for 25, 29 and 33 are :\br
\def\bbb{~~~~~~~~}
[-8 -9 22 23 18 19 -12 -13 -2 -3 \bbb\bbb          17 16 5  4 -24 -25   15 14        \bbb\bbb   -20 -21  7  6 -10 -11  1]\br
[-8 -9 26 27 22 23 -12 -13 -2 -3 \bbb      21 20   17 16 5  4 -28 -29   15 14   19 18    \bbb   -24 -25  7  6 -10 -11  1]\br
[-8 -9 30 31 26 27 -12 -13 -2 -3   25 24   21 20   17 16 5  4 -32 -33   15 14   19 18   23 22   -28 -29  7  6 -10 -11  1]
\halign{#&~#&~#&~#$\ldots$\cr
(25 10 20  4 13 17 10 24 12 \hss        15 ~9 &4 11 17 20 25 16  4 12 24 10  4 20 \hss 6&       \hss&13  9\cr
(29 14 24  4 13 17 10 28 16 \hss   8 10 19 11 &6 13 21 24 29 20  6 16 28 10  4 22 \hss 8&24 22  \hss&13  9\cr
(33 18 28  4 13 17 10 32 20 12 14 10 12 23 13 &8 15 25 28 33 24  8 20 32 10  4 24 \hss10&26 24 28 26&13  9\cr
}\halign{$\ldots$#&&~#\cr
25 18 15 \hss 7& 4 10   \hss22&10 24\hss    6 20  4 10)\cr
29 22 19 \hss11& 4 14  6\hss26&10 28\hss10  6 24  4 10)\cr
33 26 23 \hss15& 4 18  6 10 30&10 32 14 10  6 28  4 10)\cr}
This provides odd generalized fortuitous sequences for all $n\ge23$.
Two generalized fortuitous sequences seem to combine only if the first one is a regular fortuitous sequence.

\titre{VI Stack of 24 pancakes}
There is no even fortuitous sequence for $n=24$. But
an exhaustive search of all flip sequences increasing potential function by 4 at each step, succeeds in 7.5 hours and yields
$S_{24}$=(24 18  6 20 11 18 23 21 10 13  6 17 22  7 14 10  5 17 19  8  5 15  5 20 12 10 24 18 11 15  9 23 16  5 19  7  9)
proving thereby that $g(-I_{24})=37$.

\titr{VIa Stacks of 22 and 21 pancakes}
A similar search for $n=22$ fails in 88 minutes proving thereby that $g(-I_{22})>34$.
An exhaustive search of sequences of length 35 is possible but trickier.
The potential function may increase by less than 4 at some steps. So inner singletons may appear.
When an inner singleton appears or when it disappears, the potential function increases by at most 3.
So the potential function is improved if inner singletons have potential -1 instead of 0.
Now it increases by at most 2 when a singleton appears and by at most 4 when it disappears.
In other words an inner singleton is seen as a block of length $l=1$ and of potential $3l-4=-1$.
We may also see an inner singleton as a clan of length $l=1$ and of potential $4-3l=1$.
So we may consider a stack of pancakes as a stack of clans and blocks of undefined lengths.
A flip may split a clan into two smaller clans. It may glue two blocks into one bigger block.
At any time an inner clan may turn into a block. Its length is forced to 1, and its potential varies from 1 to -1.
Meanwhile the potential of the stack decreases by 2.
So a flip may glue a clan and a block or two clans yielding a block.
The lengths of the clans are unknown. The lengths of the blocks can be computed along. But we may also ignore them,
if we keep only the sum of their lengths (or the number of adjacencies).
An exhaustive search dealing with such stacks of clans and blocks of undefined lengths yields in 27 minutes sequence
$S_{21}=($21 9 13 11 20 5 3 16 4 8 19 13 21 15 3 5 8 20 10 4 18 5 3 11 15 11 19 16 3 5 11 13 9) of length 33,
proving thereby that $g(-I_{21})=33$.

Since $g(-I_{21})=33$, $g(-I_{22})>34$ and $g(-I_{n+1})\le g(-I_n)+2$ we conclude that $g(-I_{22})=35$.

\titr{VIb Stack of 20 pancakes}
Similarly an exhaustive search for an optimal sequence of length 31 sorting $-I_{20}$ fails,
proving that $g(-I_{20})=32$ since $g(-I_{19})=30$.

\titre{VII Results}
We have produced for every integer $n\in\{15,19,23\}$ or $n\ge25$,
an explicit patchwork yielding an even fortuitous sequence or a generalized odd fortuitous sequence (according to parity of $n$),
proving thereby that for all these numbers $g(-I_n)=\left\lceil{3\over2}n+1\right\rceil$.
We can achieve the same result in another way.
We have fortuitous sequences for $n\in\{15,19,23,25,29,33,26,28,30,32,34,36,38,40,42,44,46,48\}$
produced as palindromic fortuitous sequences when $n\equiv3\pmod4$ or
as triple fortuitous sequences for odd $n\ge29$ or
as even fortuitous sequences for even $n$.
Each of these sequences can be checked.
We can then complete this set of fortuitous sequences by combining every sequence with the sequence for 15 or 26 according to its parity.
So we get fortuitous sequences for all $n\ge26$.
With $S_{21}$ and $S_{24}$ and the two generalized fortuitous sequences for 19 and 25
and fortuitous sequences (2 1$)^2$ and (3 2$)^3$, we have proved that $g(-I_n)=\left\lceil 3n/2\right\rceil{+}1$
for all $n\in\{2,3,15,19,21\}$ and all $n\ge23$.
Values of $g(-I_n)$ have been given by Cohen and Blum for $n\le18$ and
by Cibulka for $n=20$ and for $n\equiv3\pmod4,~n\ge15$.
\halign{\hss$#$\hss~&&~\hss#\cr
n                                      & 1 & 2 &  3 &  4 &  5 &  6 &  7 &  8 &  9 & 10 & 11 & 12 & 13 & 14 & 15 & 16 & 17 & 18 & 19 & 20 & 21 & 22 & 23 & 24 & 25 & 26 & 27 \cr
g(-I_n)                                & 1 & 4 &  6 &  8 & 10 & 12 & 14 & 15 & 17 & 18 & 19 & 21 & 22 & 23 & 24 & 26 & 28 & 29 & 30 & 32 & 33 & 35 & 36 & 37 & 39 & 40 & 42 \cr
\left\lceil{3\over 2}n\right\rceil{+}1 & 3 & 4 &  6 &  7 &  9 & 10 & 12 & 13 & 15 & 16 & 18 & 19 & 21 & 22 & 24 & 25 & 27 & 28 & 30 & 31 & 33 & 34 & 36 & 37 & 39 & 40 & 42 \cr
\Delta                                 &-2 &   &    &  1 &  1 &  2 &  2 &  2 &  2 &  2 &  1 &  2 &  1 &  1 &    &  1 &  1 &  1 &    &  1 &    &  1 &    &    &    &    &    \cr}

All the fortuitous sequences shown in this paper but $S_{21}$ may be computed by the program in gnu C of annex.
To get the computational times given here, it was compiled with gcc 4.8.3 and run on an AMD FX-4130.
E.g. {\tt triple(37,0)} finds and prints 24 triple fortuitous sequences in 10.5s for \n=37 but
{\tt triple(39,1)} finds and prints 5 triple palindromic fortuitous sequences at once for \n=39.
As well {\tt palin(23)} finds and prints 4 palindromic fortuitous sequences in 0.008s for \n=23 and
{\tt fortui(n,sym)} looks for one (if {\tt ALL}=0) or all generalized odd fortuitous sequences or even fortuitous sequences according to parity of $n$.
If {\tt sym}=1 it adds condition $f=-f^{-1}$ and computes only double sequences.
If {\tt sym}=2 it adds condition $f=f_n\circ -f\circ f_n$ and computes only even palindromic sequences centered on $n$.
If {\tt sym}=3 it adds both conditions and computes only palindromic double sequences.
If {\tt sym}=4 it adds condition $f=f_n\circ f^{-1}\circ f_n$ and computes only palindromic sequences centered on $n{-}1$.
If {\tt sym}=0 it adds conditions depending on $n\bmod4$, such as $f(1)=-8$ if $n$ is odd, and produces at once an imposed fortuitous sequence.
The program combines also small fortuitous sequences to produce and print a fortuitous sequence
for every $n$ from 26 to 199, and checks that the produced sequences are fortuitous.
In {\tt main()} computations and printings may be disabled by {\tt if(0)} or enabled by {\tt if(1)}.
This program is provided in file {\tt concat.c} with this document.

\titre{VIII Bibliography}

[1] William H. Gates, Christos H. Papadimitriou, Bounds for sorting by prefix reversal, Discrete Mathematics, Volume 27, Issue 1, 1979, Pages 47-57, ISSN 0012-365X

[2] David S. Cohen, Manuel Blum, On the problem of sorting burnt pancakes, Discrete Applied Mathematics, Volume 61, Issue 2, 28 July 1995, Pages 105-120, ISSN 0166-218X

[3] Mohammad H. Heydari and I. Hal Sudborough. 1997. On the diameter of the pancake network. J. Algorithms 25, 1 (October 1997), 67-94

[4] Josef Cibulka, On average and highest number of flips in pancake sorting, Theoretical Computer Science, Volume 412, Issues 8-10, 4 March 2011, Pages 822-834, ISSN 0304-3975
\titre{IX Annex : program in gnu C}
\verbatimcinput{concat.c}
\bye